# Hybrid Indexes to Expedite Spatial-Visual Search

Abdullah Alfarrarjeh, Cyrus Shahabi
Integrated Media Systems Center, University of Southern California, Los Angeles, CA 90089
{alfarrar,shahabi}@usc.edu

## ABSTRACT
Due to the growth of geo-tagged images, recent web and mobile applications provide search capabilities for images that are similar to a given query image and simultaneously within a given geographical area. In this paper, we focus on designing index structures to expedite these *spatial-visual* searches. We start by baseline indexes that are straightforward extensions of the current popular spatial (R*-tree) and visual (LSH) index structures. Subsequently, we propose hybrid index structures that evaluate both spatial and visual features in tandem. The unique challenge of this type of query is that there are inaccuracies in both spatial and visual features. Therefore, different traversals of the index structures may produce different images as output, some of which more relevant to the query than the others. We compare our hybrid structures with a set of baseline indexes in both performance and result accuracy using three real world datasets from Flickr, Google Street View, and GeoUGV.

## 1. INTRODUCTION

Due to the ubiquity of smartphones equipped with high-resolution cameras and location sensors (GPS, compass, and gyroscope units), users can conveniently take a photo, which is automatically tagged with its geographical location (e.g., latitude and longitude). Thus, such photos have both *visual* and *spatial* features, and we refer to them as *geo-tagged images*. Reverse geo-tagging is studied in the literature (e.g., [25]) with promising results to annotate non-geo-tagged images with spatial metadata. Consequently, massive amounts of images that are geo-tagged are being generated at an unprecedented scale. A study [7] has indicated that by the end of 2010, the number of geo-tagged images on Flickr.com reached 95 million with a weekly growth rate of around 500,000 new geo-tagged photos.

Recently, Google (https://images.google.com) and others (e.g., Yandex, https://yandex.com/images) released their own image search engines. These systems enable users to find images on the web that are similar to an example query image. Similarly, Amazon has a mobile app (called *Flow* [33]) which supports searching for products by image, and thus enables comparative shopping. These systems do not, however, support spatial search for images. On the other hand, Flickr provides location-based search services for images. In particular, Flickr's photo API [34] enables photo retrieval using their geo-locations. Hence, it only seems natural that the next step will be the combination of these two utilities to search images for both spatial and visual relevance. We refer to these queries as *spatial-visual queries*. In fact, Google's Goggles mobile app (www.google.com/mobile/goggles) already supports searching by an example image around the user location (i.e., spatial-visual search). Such mobile and web apps have utilities in many application domains such as in online shopping, tourism, entertainment, and for searching personal photo galleries on the cloud.

The large scale of these geo-tagged image datasets and the demand for real-time response make it critical to develop efficient spatial-visual query processing mechanisms. Towards this end, we focus on designing index structures that expedite the evaluation of spatial-visual queries. This problem has some similarities to spatial-textual indexing and query processing techniques which have been studied in the past [10, 11, 13, 20, 21]). In our problem, however, text is replaced with images, resulting in two major distinctions. First, typically compared to text, an image has many more dimensions (i.e., feature points represented by high dimensional vectors) which requires the utilization of high dimensional index structures. Hence, a completely different index structure (as compared to inverted files for text) must be integrated with the spatial indexes. Second, retrieving relevant images is challenging due to two types of inaccuracies: spatial and visual. Spatial inaccuracy comes from the fact that the geo-location of the image usually reflects the camera location but not the exact location of the taken image (i.e., the scene captured by the image itself). Hence, for some images even though their locations are outside the spatial range, the images might still capture areas inside the spatial query range. Alternatively, for some images whose locations are inside the spatial query range, the images capture scenes outside the range. The visual inaccuracy comes from the fact that typical visual index structures use dimensionality reduction techniques to expedite the search in lower dimension. Thus, an image may not be considered a match in its reduced dimension but still relevant (i.e., false negatives). Due to these inaccuracies, it is not sufficient to simply find the matched results of a spatial-visual query but we need to explore the search space and the results variously. Therefore, the study of spatial-visual search is different from the study of spatial-textual search and has its own unique challenges.

For geo-tagged image datasets, the first intuitive indexing approach is to create two separate indexes, one for spatial data and another for visual data. In particular, we use R*-tree for indexing spatial data and locality sensitive hashing (LSH) index for indexing the visual features. Another naïve approach is to organize the dataset using one of its data types, either spatial or visual, using one of the popular index structures and then augment that data structure with the features of the other type. We study both variations of this approach: augmented R*-tree or augmented LSH. In addition to these three baseline approaches, we propose a set of hybrid index structures based on R*-tree and LSH. Our hybrid approaches are basically two-level index structures consisting of one primary index structure associated with a set of secondary structures. Thus, there are again two variations: using R*-tree as a primary structure (termed *Spatial First Index*) or alternatively using a primary LSH (termed *Visual First Index*). Moreover, instead of using basic secondary structures in these two-level structures, another variation is to augment the secondary structures with additional features which are indexed in the primary structure. This approach yields two other variations: *Augmented Spatial First Index* and *Augmented Visual First Index*. In this paper, we study and evaluate all these variations.

To the best of our knowledge we are the first to propose and study empirically a class of hybrid index structures for spatial-visual search queries. We evaluate and compare our proposed index structures to a set of baselines in terms of performance and accuracy using three real world geo-tagged image datasets (Flickr [27], Google Street View [25], and GeoUGV [26]). We also propose a cost model in terms of space and query time to evaluate our index structures. The experimental results showed that all hybrid structures outperformed the baselines with a maximum speed-up factor of 42.7. The choice between the hybrid variants (i.e., spatial first or visual first) depends mainly on query selectivity. The hybrid structures cannot always achieve 100% recall due to their utilization of LSH, which is an information retrieval index structure, while one of the baselines that is not LSH-based, can achieve higher recall at the cost of sacrificing performance.

The remainder of this paper is organized as follows. Section 2 introduces a set of preliminary definitions, the problem statement, and background about the state of the art techniques for spatial and visual indexing. Section 3 introduces a set of index structures for geo-tagged images. Section 4 reports on our experimental results. In Section 5, we review the related work. Finally, in Section 6, we conclude and discuss future work.

## 2. PRELIMINARIES
## 2.1 Geo-Tagged Image Model
Each image is represented by two vectors: *spatial* and *visual*.

### 2.1.1 Spatial Vector
Each image $I$ is tagged with a spatial location $I.s$. Without loss of generality (WLOG) and to simplify illustrations, we assume the location is represented by a 2-dimensional point composed of latitude and longitude values[1].

### 2.1.2 Visual Vector
Image content can be represented by various features including; color (e.g., color histogram [3]), texture (e.g., Gabor [3]), local interest points (e.g., SIFT [2]), and bag-of-words descriptor [1]. Our visual representation model is based on the state-of-the-art deep convolutional neural networks (CNNs) [16]. In general, CNN is a hierarchical architecture which consists of convolutional and sub-sampling layers followed by fully connected layers. CNN was first proposed and used in computer vision (e.g., image classification [16]). Recent studies (e.g., [14]) show that CNN is very promising in image representation for performing content-based image retrieval.

CNN utilizes a pre-trained model for extracting a rich image feature vector consisted of 4096 dimensions. Due to high dimensionality, dimension reduction techniques (e.g., principal component analysis (PCA)) can be used to generate compact representation by selecting the most important internal components (i.e., dimensions) of each vector. It was shown experimentally [18] that CNN feature vectors can be considerably reduced in dimension without significantly degrading retrieval quality. In this paper, WLOG, we represent the visual vector of an image $I.v$ as a 150-dimensional vector derived from its 4096-dimensional CNN vector.

## 2.2 Background
### 2.2.1 R*-tree
The most popular index structure for spatial data is R-tree. An R-tree [5] is a multiway tree which constructs a recursive hierarchical cover of the data space. Each node in the tree corresponds to the smallest d-dimensional rectangle that encloses its child nodes. In R-tree, spatial cover of a node might overlap other nodes which make answering a query more expensive. R*-tree (a revised version of R-Tree) [6] uses sophisticated techniques to minimize the overlap between nodes by revising the R-tree node split algorithm to consider a combination of parameters to minimize node area, margin and overlap.

Given a spatial range, the root of R*-tree is checked to recursively query the child nodes whose minimum bounding rectangles (MBRs) overlap with the query range. Once finding a leaf node overlapping with the query range, all of its geo-points are examined to retrieve the points inside the query range. In this paper, we use R*-tree for indexing the spatial feature vector ($I.s$) of an image[2].

### 2.2.2 Locality Sensitive Hashing (LSH)
Given that images are represented by feature-rich vectors, finding relevant images is defined as similarity search in a high-dimensional space. Several tree-based index structures (e.g., M-Tree [15] and KD-Tree [17]) have been proposed for exact-result similarity search with low-dimensional space. Meanwhile, the performance of these index structures in high-dimensional space degrades to less than that of the linear scan approaches [24]. To perform approximate similarity search, several methods have been proposed (e.g., [31] [32]), among which, locality-sensitive hashing (LSH) [32] is widely used for its theoretical guarantees and empirical performance. The key notion of locality sensitive hashing (LSH) is to use a set of hash functions, from a hash family $\mathcal{H}$ in a metric space $\mathcal{M}$, which map similar objects into the same buckets with higher probabilities than dissimilar objects. Given a particular metric $\mathcal{M}$ and corresponding hash family $\mathcal{H}$, LSH index structure maintains a number of hash tables containing references to the objects in the dataset.

Indyk et al. [32] originally introduced LSH hash families suitable only for Hamming space and later other variations of LSH hash families have been devised for different metric spaces [19]. In this paper, we assume the metric space is the d-dimensional Euclidean space $\mathbb{R}^d$, which is the most commonly used metric space. In Euclidean space, Datar et al [24] defined an LSH family as follows:

$$\mathcal{H}(o) = \langle h_1(o), h_2(o), \ldots, h_F(o) \rangle$$

$$h_i(o) = \left\lfloor \frac{\vec{a}_i \cdot \vec{o} + b_i}{W} \right\rfloor, 1 \leq i \leq F$$

Given an object $\vec{o} \in \mathbb{R}^d$, the function first projects the object onto a random vector $\vec{a}_i \in \mathbb{R}^d$ whose entries are chosen independently from the standard normal distribution $\mathbb{N}(0, 1)$. Subsequently, the projected vector is shifted by a real number $b_i$ drawn from the uniform distribution $[0, W)$ where $W$ is a user specified constant. For each $i$, $\vec{a}_i$ and $b_i$ are sampled independently. With LSH, the parameters $F$ and $W$ control the locality sensitivity of the hash table. The index data structure consists of $T$ hash tables with independent $F$ hash functions. For reducing the number of hash

---

[1] For example, one can represent the spatial feature of an image with a spatial extent, such as its Field-of-View (FOV) as in [28].

[2] If the spatial feature vector is not a point, then other index structures may become more appropriate, for example, for FOV's, one may want to use OR-trees [28].

tables in LSH, the multi-probe LSH algorithm is proposed [30] which basically probes several buckets in the same hash table. In addition, for predicting the values of the parameters *F* and *W*, Dong et al [4] proposed a statistical model which predicts the values given a small sample dataset.

Given a query image $I_Q$, a visual feature vector ($I_Q.v$) is computed to be projected and hashed to find the buckets which contain the images with the highest probability to be similar. The stored images in these candidate buckets are considered as relevant images. Subsequently, a sub-list of relevant images whose distance to the query image is less than a similarity threshold is retrieved. Because LSH uses the projection operation for dimensionality reduction, querying LSH results in a best-match list of relevant images which is not necessary to be all relevant images. In particular, there might be a visual vector $I_j.v$ that is hashed to a bucket other than the bucket assigned to $I_Q.v$ however their visual similarity distance is less than the query similarity threshold so $I_j.v$ is considered as false miss.

## 2.3 Terminologies

**DEFINITION 1 (Geo-tagged Image Dataset):** We have a dataset of *n* geo-tagged images $D = \{I_0, I_1... I_{n-1}\}$ that is stored in the disk. Each geo-tagged image *I* is represented by a pair of spatial (*I.s*) and visual (*I.v*) vectors.

**DEFINITION 2 (Spatial-Visual Query):** It is defined as $Q = (Q.s, Q.v)$, where *Q.s* is the spatial part of the query *Q* (e.g., spatial range or nearest neighbor) and *Q.v* is the visual part (e.g., top N similar images or similar images to an example image within certain similarity threshold).

**Problem Statement:** Given a geo-tagged image dataset *D*, we would like to create a disk-resident index structure which expedites spatial-visual query *Q*.

In this paper, we propose a set of index structures which can be utilized for any type of spatial-visual queries, but throughout the discussion we focus only on one type which is a spatial-visual range query $Q_{range}$. With $Q_{range}$, the spatial part (*Q.s*) is represented by a 2D orthogonal range (i.e., bottom-left coordinates are (min(x), min(y)), and top-right coordinates are (max(x), max(y))), while the visual part *Q.v* is a visual range represented by an image $I_Q$ and a visual similarity threshold $\sigma$. Given that $I_Q.v \in \mathbb{R}^d$, the visual similarity distance $\Phi(I_k.v, I_Q.v)$ between $I_Q.v$ and a visual vector of an image in the dataset $I_k.v$ is defined using the Euclidean distance:

$$\Phi(I_k.v, I_Q.v) = \sqrt{\sum_{j=0}^{d}(I_k.v_j - I_Q.v_j)^2}$$

$I_k$ is considered as an output of a spatial-visual range query $Q_{range}$ if and only if $I_k.s$ is inside both *Q.s* and *Q.v*.

Due to the inaccuracies associated with the spatial-visual search, we categorize the results of spatial-visual range query $Q_{range}$ into the following three classes.

**DEFINITION 3 (Spatially-Visually Matched and Relevant Image** (*SV-Match-Rel*)**):** An image $I_k$ is spatially-visually matched if for a given $Q_{range}$, its spatial vector $I_k.s$ is inside *Q.s* and the image itself is similar to the query image $I_Q$ (i.e., $\Phi(I_k.v, I_Q.v) \leq \sigma$). Thus, $Q_{range}$ result includes $I_k$.

**DEFINITION 4 (Spatially Unmatched but Relevant Image** (*S-UNMatch-Rel*)**):** An image $I_k$ is spatially unmatched but relevant if for a given $Q_{range}$, its spatial vector $I_k.s$ (i.e., the camera location of the image) is outside *Q.s* but the image itself is similar to the query image $I_Q$ and satisfies the visual similarity distance threshold (i.e., $\Phi(I_k.v, I_Q.v) \leq \sigma$). $Q_{range}$ result does not include $I_k$ but obviously $I_k$ is relevant.

**DEFINITION 5 (Visually Unmatched but Relevant Image** (*V-UNMatch-Rel*)**):** An image $I_k$ is visually unmatched but relevant if for a given $Q_{range}$, its spatial vector $I_k.s$ is inside *Q.s* but it is not reported by LSH as visually match; however, the image itself is similar to the query image $I_Q$ and satisfy the visual similarity distance threshold (i.e., $\Phi(I_k.v, I_Q.v) \leq \sigma$). This happens due dimension reduction of LSH where the matched images in the reduced dimensions are only a subset of all relevant images. Here, $Q_{range}$ result does not include $I_k$ but obviously $I_k$ is relevant.

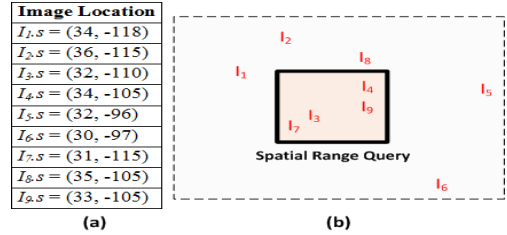

**Figure 1: (a) Locations of Set of Geo-tagged Images (b) Spatial Range Query relative to the image locations**

**Table 1: Visual Similarity Distance between the Sample Geo-tagged Images and the Query Image ($I_Q$.v)**

| Similarity Distance to the Query Image ($I_Q$.v) | | |
|---|---|---|
| $\Phi(I_1.v, I_Q.v) = 1.5$ | $\Phi(I_4.v, I_Q.v) = 0.3$ | $\Phi(I_7.v, I_Q.v) = 0.6$ |
| $\Phi(I_2.v, I_Q.v) = 0.6$ | $\Phi(I_5.v, I_Q.v) = 0.2$ | $\Phi(I_8.v, I_Q.v) = 0.4$ |
| $\Phi(I_3.v, I_Q.v) = 0.1$ | $\Phi(I_6.v, I_Q.v) = 0.8$ | $\Phi(I_9.v, I_Q.v) = 0.4$ |

**Table 2: Notation Table**

| | |
|---|---|
| $M$ | The number of leaf nodes in a primary R*-Tree |
| $\overline{M}$ | The average number of leaf nodes in a secondary R*-Tree |
| $B$ | The maximum number of buckets among all hash tables in a primary LSH |
| $\overline{B}$ | The average number of buckets in a secondary LSH |
| $C(b)$ | The size of bucket *b* in LSH |
| $m_b$ | A leaf node *m* in a secondary R*-tree attached to a bucket *b* belonging to a primary LSH |
| $b_m$ | A bucket *b* in a secondary LSH attached to a leaf node *m* belonging to a primary R*-tree |
| $P_s(X)$ | The size of spatial data referenced by the entity x (i.e., leaf node or bucket) |
| $P_v(X)$ | The size of visual data referenced by the entity x (i.e., leaf node or bucket) |
| $P_{disk}$ | The size of a page disk |
| $T_{disk}$ | The time cost of one disk access |

## 3. INDEXING APPROACHES

In this section, we first discuss a set of baselines and then our hybrid indexes in terms of their structures, query execution, accuracy of query results and performance. We also present the space and query time cost for each structure. The notations used in the cost model description are listed in Table 2. Eq. (1) represents the space cost model which is the storage sum of both index entities and data, while Eq. (2) represents the query time model in terms of I/O operations which is the sum of the time for loading the index entities and the data referenced by them. Table 3 and 4 summarizes the components of both Eq. (1) and (2) amongst various index structures, respectively.

$$SpaceCost = S_R + S_{LSH} + S_{Data} \qquad (1)$$

$$QueryIOCost = T_{disk} * (T_R + T_{LSH} + T_{Data}) \quad (2)$$

Throughout this section we use the following running example:

*Example: Suppose a geo-tagged image dataset includes the nine images whose geographic coordinates are listed in Fig. 1-a. Consider a spatial-visual range query $Q_{range}$, where its spatial range Q.s (as shown in Fig. 1-b) is defined by a rectangle whose geographic coordinates of the minimum point (bottom-left corner) are (30, -116) and the coordinates of the maximum point (top-right) are (34,-104). The visual range part Q.v is composed of an example image $I_Q$ and a similarity threshold $\sigma = 0.5$. The visual similarity distances between $I_Q$ and each of the nine images $I_k$ (i.e., $\Phi(I_k.v, I_Q.v)$) are shown in Table 1.*

## 3.1 Baseline Index Structures

### 3.1.1 Double Index (DI)

This structure is composed of two separate index structures: R*-tree (Fig. 2) and LSH (Fig.3). For each image $I$, its spatial vector $I.s$ is indexed by R*-tree and its visual vector $I.v$ by LSH, independently. In Fig. 2, R*-tree contains two internal nodes ($R_a$, $R_b$). $R_a$ contains the leaf nodes $R_1$, $R_2$ and $R_3$ while $R_b$ contains $R_4$, $R_5$, and $R_6$. Each leaf node contains a list of geo-tagged images, but for simplicity we only show our nine sample images. Meanwhile, LSH is composed of two hash tables as shown in Fig. 3. The first hash table contains four buckets (i.e., $B_1$ to $B_4$) and the second table contains three buckets $B_5$ to $B_7$. For simplicity, the figure shows only the distribution of the nine images amongst the buckets. In reference to Eq. (1), the space cost with *DI* is the sum of the storage of the R*-tree index (i.e., $S_R$), the LSH index (i.e., $S_{LSH}$), and the image dataset (i.e., $S_{Data}$). The storage of R*-Tree which has $x$ leaf nodes is $O(x)$ [12] because each node represents one disk page. With LSH, there is no limit on the size of a bucket so the storage of LSH is the total size of all buckets $B$ divided by the disk page size $P_{disk}$. Table 3 shows the space cost components of *DI*.

At query time, R*-tree is queried using the spatial part of the query $Q.s$ and LSH is queried using $Q.v$. The intermediate results retrieved from R*-tree satisfy only the spatial part while the intermediate results retrieved from LSH satisfy only the visual part. To answer the spatial-visual query, an intersection filter is executed on the intermediate results[3].

With our running example, as shown in Fig 2 the spatial range $Q.s$ intersects the MBRs of the leaf nodes $<R_3, R_4>$. Hence, R*-tree loads these leaf nodes in addition to two internal nodes ($R_a$, $R_b$) and retrieves the spatial vectors of the images $<I_3, I_4, I_7, I_8, I_9>$. R*-tree discards $I_8$ because it is outside the spatial range query and outputs the list $<I_3, I_4, I_7, I_9>$ as intermediate results. Meanwhile, to execute the visual range query $Q.v$, the visual vector of the query image $I_Q.v$ is hashed using LSH. Suppose that $I_Q.v$ is hashed to the buckets $<B_3, B_6>$, then LSH loads these two buckets from the disk. These two buckets contain the candidate list $<I_3, I_4, I_5, I_6, I_7, I_8>$. Based on the visual similarity distance of our images $I_k.v$ from $I_Q.v$ in Table 1, LSH reports $<I_3, I_4, I_5, I_8>$ as intermediate result because their distance is less than $\sigma$ (i.e., $\Phi(I_k.v, I_Q.v) \leq 0.5$). After applying the intersection filter on the individual results from R*-tree and LSH, the list $<I_3, I_4>$ is reported as the final result.

---

[3] Note that for kNN queries, more complex rank-merge algorithms, such as Fagin's [29], can be used instead.

***Result Accuracy:*** *Double Index* can only retrieve a subset of all relevant images. This is because while it can retrieve *SV-Match-Rel* images, it fails to retrieve *S-UNMatch-Rel* and *V-UNMatch-Rel* ones. *S-UNMatch-Rel* images are missed because of treating the spatial filter strictly while *V-UNMatch-Rel* images are missed because of the LSH inaccuracy caused by its dimension reduction.

***Index Performance:*** The main disadvantage of this technique is that the intersection filter can be expensive if the size of the intermediate results is large. The extreme case is when the intermediate results do not overlap at all, in which case each index is used to retrieve a set of "useless" intermediate results. In reference to Eq. (2), the query cost with *DI* is the sum of the time for loading the nodes of R*-tree (i.e., $T_R$), buckets of LSH (i.e., $T_{LSH}$), and data referenced by both R*-tree leaf nodes and LSH buckets (i.e., $T_D$). In the worst case, having a query intersecting with all leaf nodes requires loading all leaf nodes to retrieve the results. Hence, the cost of querying R*-tree which has $x$ leaf nodes is the time to load $O(x)$ leaves. Meanwhile, LSH requires loading only the buckets to which $I_Q.v$ is hashed (i.e., $O(C(b))/P_{disk}$). Table 4 shows the query cost components of *DI*.

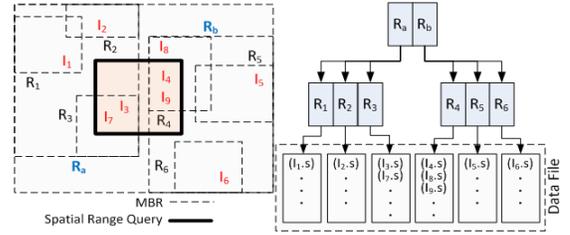

**Figure 2: Double Index Structure - R*-tree**

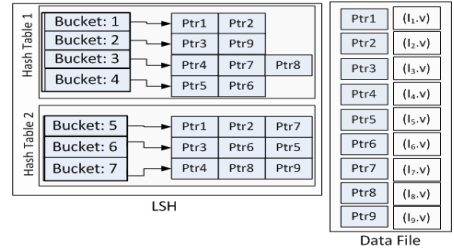

**Figure 3: Double Index Structure - LSH**

### 3.1.2 Augmented R*-tree (Aug R*-tree)

This approach organizes data spatially using an augmented version of R*-tree whose leaf nodes store not only pointers to spatial vectors ($I.s$) of the geo-tagged images but also pointers to their visual vectors ($I.v$). When building the index structure, R*-tree uses only the spatial part ($I.s$) of the images to distribute objects across its nodes. The space cost with *Aug R*-tree* is shown in Table 3 where it is similar to that with *DI* but discards $S_{LSH}$.

To execute a spatial-visual query, first the spatial part of the query $Q.s$ is utilized to retrieve an intermediate result list of images which spatially satisfies the query. Subsequently, the intermediate results are inspected sequentially to discard all images that are irrelevant to the visual part of the query $Q.v$. Therefore, the final query result satisfies the visual filter as well.

With our running example, querying *Aug R*-tree* requires loading the leaf nodes $<R_3, R_4>$ and it returns $<I_3, I_4, I_7, I_9>$ as intermediate result. Later on, the filtering step discards $<I_7>$ because its visual similarity distance to $I_Q$ is greater than $\sigma$ (i.e., $\Phi(I_7.v, I_Q.v) > 0.5$). Hence, the final result is $<I_3, I_4, I_9>$.

**Table 3: Space Cost of Various Index Structures**

|  | $S_R$ | $S_{LSH}$ | $S_{Data}$ |
|---|---|---|---|
| DI | $O(M)$ | $O(\sum_{b=1}^{B} C(b)) / P_{disk}$ | $O(\sum_{m=1}^{M} P_S(m) + \sum_{b=1}^{B} P_V(b)) / P_{disk}$ |
| Aug R*-tree | $O(M)$ | - | $O(\sum_{m=1}^{M} P_S(m) + \sum_{m=1}^{M} P_V(m)) / P_{disk}$ |
| Aug LSH | - | $O(\sum_{b=1}^{B} C(b)) / P_{disk}$ | $O(\sum_{b=1}^{B} P_s(b) + \sum_{b=1}^{B} P_v(b)) / P_{disk}$ |
| SFI | $O(M)$ | $O(\sum_{m=1}^{M} \sum_{b=1}^{\bar{B}} C(b_m)) / P_{disk}$ | $O(\sum_{m=1}^{M} P_S(m) + \sum_{m=1}^{M} \sum_{b=1}^{\bar{B}} P_V(b_m)) / P_{disk}$ |
| VFI | $O(\sum_{b=1}^{B} O(\bar{M}_b))$ | $O(\sum_{b=1}^{B} C(b)) / P_{disk}$ | $O(\sum_{b=1}^{B} \sum_{m=1}^{\bar{M}} P_S(m_b) + \sum_{b=1}^{B} P_V(b)) / P_{disk}$ |
| Aug SFI | $O(M)$ | $O(\sum_{m=1}^{M} \sum_{b=1}^{\bar{B}} C(b_m)) / P_{disk}$ | $O(\sum_{m=1}^{M} P_S(m) + \sum_{m=1}^{M} \sum_{b=1}^{\bar{B}} P_V(b_m)) / P_{disk}$ |
| Aug VFI | $O(\sum_{b=1}^{B} O(\bar{M}_b))$ | $O(1)$ | $O(\sum_{b=1}^{B} \sum_{m=1}^{\bar{M}} P_S(m_b) + \sum_{b=1}^{B} P_V(b)) / P_{disk}$ |

**Table 4: Query I/O Cost of Various Index Structures**

|  | $T_R$ | $T_{LSH}$ | $T_{Data}$ |
|---|---|---|---|
| DI | $O(M)$ | $O(C(b))/P_{disk}, b \in [1,B]$ | $(O(\sum_{m=1}^{M} P_S(m) + P_V(b))/P_{disk}, b \in [1,B]$ |
| Aug R*-tree | $O(M)$ | - | $O(\sum_{m=1}^{M} P_S(m) + \sum_{m=1}^{M} P_V(m))/P_{disk}$ |
| Aug LSH | - | $O(C(b))/P_{disk}, b \in [1,B]$ | $O(\sum_{b=1}^{B} P_s(b) + \sum_{b=1}^{B} P_v(b))/P_{disk}$ |
| SFI | $O(M)$ | $\sum_{m=1}^{M} O(C(b_m))/P_{disk}, b \in [1,\bar{B}]$ | $O(\sum_{m=1}^{M} P_S(m) + \sum_{m=1}^{M} P_V(b_m))/P_{disk}, b \in [1,\bar{B}]$ |
| VFI | $O(\bar{M}_b), b \in [1,B]$ | $O(\sum_{m=1}^{\bar{M}} P_S(m_b) + P_V(b)/P_{disk}), b \in [1,B]$ | $O(\sum_{b=1}^{B} \sum_{m=1}^{\bar{M}} P_S(m_b) + \sum_{b=1}^{B} P_V(b))/P_{disk}$ |
| Aug SFI | $O(\log M)$ | $\sum_{m=1}^{M} O(C(b_m))/P_{disk}, b \in [1,\bar{B}]$ | $O(\sum_{m=1}^{M} P_S(b_m) + \sum_{m=1}^{M} P_V(b_m))/P_{disk}, b \in [1,\bar{B}]$ |
| Aug VFI | $O(\bar{M}_b), b \in [1,B]$ | $O(1)$ | $O(\sum_{m=1}^{\bar{M}} P_S(m_b) + \sum_{m=1}^{\bar{M}} P_v(m_b)), b \in [1,B]$ |

***Result Accuracy:*** As with *DI*, *Aug R*-tree* can retrieve *SV-Match-Rel* images but it fails to retrieve *S-UNMatch-Rel* ones. However, unlike *DI*, it can retrieve *V-UNMatch-Rel* images because it stores the visual vectors $I_k.v$ without any form of dimension reduction.

***Index Performance:*** The main drawback of this index structure is that it only considers the spatial vectors *I.s* to organize the image dataset in the structure. Consequently, if the spatial selectivity of $Q_{range}$ is low, then the performance of $Q_{range}$ deteriorates because it will retrieve a large number of images that satisfy the spatial part (i.e., *Q.s*) but may be discarded later in the visual filtering step. In the worst case (i.e., when none of the images satisfy *Q.v*), the entire result set of the spatial query will be discarded after their retrieval from the disk. As shown in Table 4, the query cost with *Aug R*-tree* is similar to that with *DI* but discards the $T_{LSH}$ part and retrieves visual data based on the size of the leaf nodes.

*3.1.3 Augmented LSH (Aug LSH)*

In contrast to *Aug R*-tree*, the augmented LSH structure organizes the geo-tagged image dataset based on its visual vectors *I.v*. However, in this modified version of LSH, the LSH buckets include pointers to both the spatial and visual vectors (*I.s*, *I.v*). The space cost with *Aug LSH* is shown in Table 3 where it is similar to that with *DI* but discards $S_R$.

At query time, LSH first retrieves all images that satisfy the visual part of the query *Q.v*. Next, it applies a spatial filter on this intermediate result to discard all images that are outside the spatial range of the query *Q.s*.

With our running example, when executing the visual query *Q.v*, LSH loads two buckets $<B_3, B_6>$ from the disk. These two buckets contain the candidate list $<I_3, I_4, I_5, I_6, I_7, I_8>$. Based on the visual similarity distance, only the sub-list $<I_3, I_4, I_5, I_8>$ is reported as the intermediate result. After applying the spatial filtering step, the final result is $<I_3, I_4>$ since neither $I_5$ nor $I_8$ satisfies the spatial query range *Q.s*.

***Result Accuracy:*** Similar to *DI*, *Aug LSH* can retrieve *SV-Match-Rel* images and it retrieves neither *S-UNMatch-Rel* nor *V-UNMatch-Rel* ones.

***Index Performance:*** The main drawback of this index structure is that, opposite to *Aug R*-tree*, it only considers the visual vectors *I.v* to organize the image dataset in the index structure. Consequently, if the visual selectivity of the query is low, then the performance of the query deteriorates because it will retrieve a large number of images that satisfy the visual query but may be discarded later in the spatial filtering step. In the worst case (i.e., when none of the images satisfy the spatial part of the query *Q.s*), the entire result set of the visual query will be discarded after retrieving them from the disk. As shown in Table 4, the query cost with *Aug LSH* is similar to that with *DI* but discards the $T_R$ part and retrieves spatial data based on the size of the buckets.

## 3.2 Hybrid Index Structures

*3.2.1 Spatial First Index (SFI)*

With this structure, first R*-tree is built on all MBRs covering the spatial scope (*I.s*) of all images. Next, all the images in each R*-tree leaf node are indexed by an LSH index based on their visual vectors *I.v*. Consequently, there is one primary R*-tree and a set of secondary LSHs corresponding to R*-tree leaf nodes (Fig. 4). The space cost with *SFI* is shown in Table 3 where the primary part (i.e., $S_R$) is similar to its peer in *DI* but in each secondary LSH the number of buckets in each hash table is $\bar{B}$ in average.

When executing a spatial-visual query, the spatial part of the query *Q.s* is used to filter out the set of R*-tree leaf nodes that contain the candidate result. If *Q.s* overlaps with multiple R*-tree leaf nodes then their corresponding LSH indexes are queried using *Q.v*. Hence, this may generate multiple LSH sub-results which require to be merged to obtain the *Q.v* result. To retrieve the result of a spatial-visual query, R*-tree not only reports the candidate leaf nodes but also the candidate spatial vectors contained in those nodes. Hence, R*-tree should retrieve the spatial vectors from the disk when loading a leaf node to examine if they satisfy *Q.s*. The visual part of the query (*Q.v*) is executed with LSHs attached to the candidate R*-tree leaf nodes. The intersection of the intermediate results retrieved from the queried LSHs and R*-tree constitute the result for the spatial-visual query.

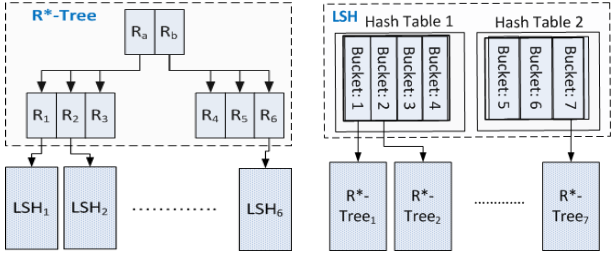

**Figure 4: Spatial First Index Structure**    **Figure 5: Visual First Index Structure**

With our running example, when executing *Q.s* the primary R*-tree loads two leaf nodes <$R_3$, $R_4$> and then retrieves their content. R*-tree reports the list <$I_3$, $I_4$, $I_7$, $I_9$> as intermediate result. The second phase is to execute *Q.v* on the secondary LSHs linked with $R_3$ and $R_4$ (referred to as $LSH_3$ and $LSH_4$, respectively). Suppose that $I_Q.v$ is hashed to the buckets $B_3$ and $B_6$ in both $LSH_3$ and $LSH_4$. Then, the list <$I_3$, $I_4$, $I_8$> is reported as another list of intermediate result. When intersecting both intermediate results, <$I_3$, $I_4$> is reported as the final result.

***Result Accuracy:*** The result of *SFI* is identical to *DI*. It retrieves *SV-Match-Rel* images but neither *S-UNMatch-Rel* nor *V-UNMatch-Rel*.

***Index Performance:*** This index structure follows a two-level data space organization where spatial data is the primary filter to prune the search space for the visual part of the query *Q.v*. Consequently, this structure performs well when the spatial selectivity of the query is high. As shown in Table 4, the query cost with *SFI* is similar to that with *DI* but the $T_{LSH}$ part is presented by querying multiple secondary LSHs. We further observe the following lemmas:

***Lemma 1:*** *QueryIOCost*(*SFI*) ≤ *QueryIOCost*(*Aug R*-tree*)

***Proof:*** Given that for each leaf node $m \in [1, M]$, a subset of data objects referenced by $m$ is also referenced by $b_m \in [1, \bar{B}]$. Hence, we can conclude that $C(b_m)/P_{disk} \leq m$ and $P_v(b_m) \leq P_v(m)$. Subsequently, we can generalize that $O(\sum_{m=1}^{M} C(b_m)/P_{disk}) = O(M)$ and $O(\sum_{m=1}^{M} P_v(b_m)) = O(\sum_{m=1}^{M} P_v(m))$. Hence, $T_{LSH}^{SFI} \leq T_R^{Aug\ R*-tree}$ and $T_{Data}^{SFI} \leq T_{Data}^{Aug\ R*-tree}$.

***Lemma 2:*** *QueryIOCost*(*SFI*) ≤ *QueryIOCost*(*DI*)

***Proof:*** Given that $b_m$ in a secondary LSH contains a sub-dataset of $b$ in the primary LSH. Subsequently, we can declare that $O(\sum_{m=1}^{M} C(b_m)/P_{disk})$, $b_m \in [1, \bar{B}] = O(C(b))$, $b \in [1, B]$ and $O(\sum_{m=1}^{M} P_s(b_m) + \sum_{m=1}^{M} P_V(b_m)) = O\ (P_s(b) + P_v(b))$. Hence, $T_{LSH}^{SFI} \leq T_{LSH}^{DI}$ and $T_{Data}^{SFI} \leq T_{Data}^{DI}$.

### 3.2.2 Visual First Index (VFI)

Opposite to the *SFI* structure in which the primary index is R*-tree, *VFI* uses LSH as its primary index in which all images are distributed into different buckets based on their visual vectors (*I.v*). As shown in Fig. 5, each LSH bucket is associated with an R*-tree to organize the images in the bucket based on their spatial vectors *I.s*. Consequently, we have one LSH and multiple secondary R*-trees based on the number of the LSH buckets. The space cost with *VFI* is shown in Table 3 where the primary part (i.e., $S_{LSH}$) is similar to its corresponding one with *DI* but in each secondary R*-tree, the number of leaf nodes is $\bar{M}$ in average.

The execution order of a spatial-visual query $Q_{range}$ with *VFI* is reversed as compared to *SFI*; the visual part *Q.v* is executed first to find the candidate buckets and then *Q.s* is executed on their corresponding R*-trees. All the visual vectors stored in the candidate buckets are retrieved to examine whether their visual similarity distance is below *σ*. Finally, the retrieved LSH result is intersected with the sub-results of the queried R*-trees.

With our running example, when executing *Q.v* the primary LSH loads two buckets <$B_3$, $B_6$> and then retrieves their contents. LSH reports the list <$I_3$, $I_4$, $I_8$> as an intermediate result. The second phase is to execute *Q.s* on the secondary R*-trees associated with $B_3$ and $B_6$ (referred to as *R*-tree$_3$* and *R*-tree$_6$*, respectively). Subsequently, the list <$I_3$, $I_4$, $I_7$> is reported as another intermediate result. When intersecting both intermediate results, <$I_3$, $I_4$> is reported as the result.

***Result Accuracy:*** The result of *VFI* is identical to *DI*.

***Index Performance:*** Since all images are indexed first by their visual vectors, this two-level index structure uses the visual part to prune the search space for the spatial part of the query *Q.s*. Consequently, this structure performs well when the visual selectivity of the query is high. As shown in Table 4, the query cost with *VFI* is similar to that with *DI* but the $T_R$ part is presented by querying multiple secondary R*-trees. We further observe the following lemmas:

***Lemma 3:*** *QueryIOCost*(*VFI*) ≤ *QueryIOCost*(*Aug LSH*)

***Proof:*** Given that for each bucket $b \in [1, B]$, a subset of data objects referenced by $b$ is also referenced by $m_b \in [1, \bar{M}]$. Hence, we can conclude that $m_b \leq C(b)/P_{disk}$ and $P_s(m_b) \leq P_s(b)$. Hence, $T_R^{VFI} \leq T_{LSH}^{Aug\ LSH}$ and $T_{Data}^{VFI} \leq T_{Data}^{Aug\ LSH}$.

***Lemma 4:*** *QueryIOCost*(*VFI*) ≤ *QueryIOCost*(*DI*)

***Proof:*** Given that $m_b$ in a secondary R*-tree contains a sub-dataset of $m$ in the primary R*-tree. Subsequently, $O(\bar{M}_b)$, $b \in [1, B] = O(M)$ and $O(\sum_{m=1}^{\bar{M}} P_s(m_b) + \sum_{m=1}^{\bar{M}} P_V(m_b)) = O(\sum_{m=1}^{M} P_s(m) + \sum_{m=1}^{M} P_V(m))$. Hence, $T_R^{VFI} \leq T_R^{DI}$ and $T_{Data}^{VFI} \leq T_{Data}^{DI}$.

Comparing the query cost with *SFI* to that with *VFI*, when $T_R^{SFI} < T_{LSH}^{VFI}$ (i.e., the spatial selectivity of the query is high and the visual selectivity of the query is low), this leads to *QueryIOCost*(*SFI*) < *QueryIOCost*(*VFI*).

## 3.3 Augmented Hybrid Index Structures

### 3.3.1 Augmented Spatial First Index (Aug SFI)

With *SFI*, to retrieve *SV-Match-Rel* images, the intermediate results from both the primary index and secondary indexes are intersected which slows down the query execution. To avoid this query performance degradation, the secondary indexes of the *SFI* structure are modified by augmenting the LSH buckets with extra pointers to *I.s*. This structure is referred to as *Augmented Spatial First Index* (*Aug SFI*). As shown in Table 3, the space cost with *Aug SFI* is identical to that with *SFI*. In this structure, maintaining additional pointers to spatial data enlarges the size of buckets marginally which is considered by $S_{LSH}$.

When executing a spatial-visual query, the spatial part *Q.s* is executed to identify the leaf nodes that overlap the spatial range query without retrieving their spatial vectors *I.s*. The secondary LSHs associated with the candidate leaf nodes are queried using *Q.v*. Since the secondary LSHs are augmented with *I.s*, they can directly discard results that do not satisfy *Q.s*. There is another variation of the query execution in which the secondary LSHs are explored to retrieve more results. When executing *Q.v.* on each secondary LSH, we can generate a set of random vectors $I_k.v$ where $\Phi(I_k.v, I_Q.v) \leq \sigma$ to query more buckets which contain potentially similar images to $I_Q.v$. This query variant (i.e., visual-explorative query; referred as *Aug SFI-E*) minimizes the effect of LSH's inaccuracy. The number of randomly generated vectors represents the visual exploration ratio (i.e., $\mathcal{E}.v$).

With our running example, when executing *Q.s* the primary R*-tree loads two leaf nodes <$R_3$, $R_4$>. Then, *Q.v* is executed on $LSH_3$ and $LSH_4$. Under the same assumption of hashing $I_Q.v$ into the buckets $B_3$ and $B_6$, the list <$I_3$, $I_4$, $I_8$> is reported as the intermediate result. Subsequently, $I_8$ is discarded because it is outside the spatial range query. Hence, <$I_3$, $I_4$> is reported as the final result.

*Result Accuracy:* The result of *Aug SFI* is identical to *SFI* and *DI*. It retrieves *SV-Match-Rel* images but fails to retrieve *S-UNMatch-Rel* and *V-UNMatch-Rel*. *Aug SFI–E* can retrieve some of *V-UNMatch-Rel* based on the exploration ratio. If each secondary LSH is extensively explored, *Aug SFI–E* can retrieve all *V-UNMatch-Rel* images

*Index Performance:* Compared to *SFI*, the query performance of *Aug SFI* is improved because it reduces the number of I/O operations. However, this structure still relies on spatial filter to prune the search space first. Consequently, *Aug SFI* performs well when the spatial selectivity of the query is high. As shown in Table 4, the query cost with *Aug SFI* is similar to that with *SFI* but minimizes the cost of $T_R$ part considerably. We further observe the following lemmas:

**Lemma 5:** *QueryIOCost(Aug SFI) ≤ QueryIOCost(SFI)*

*Proof:* with *Aug SFI* executing the query does not need to load the leaf nodes in the primary R*-tree. Given that R*-tree is a full tree, reaching the leaf nodes is bounded with its height $O(\log M)$ which is negligible because the R*-tree is a shallow tree. Hence, $T_R^{Aug\ SFI} \leq T_R^{SFI}$. In addition, for each leaf node $m \in [1, M]$, a subset of data objects referenced by $m$ is also referenced by $b_m \in [1, \bar{B}]$. Hence, we can conclude that $C(b_m)/P_{disk} \leq m$ and $P_s(b_m) \leq P_v(m)$. Subsequently, $O(\sum_{m=1}^{M} P_s(b_m)) = O(\sum_{m=1}^{M} P_s(m))$. Hence, $T_{Data}^{Aug\ SFI} \leq T_{Data}^{SFI}$.

**Lemma 6:** *QueryIOCost(Aug SFI-E) ≤ QueryIOCost(Aug R*-tree)*

*Proof:* Derived from Lemma 1 (details omitted due to space limitation).

### 3.3.2 Augmented Visual First Index (Aug VFI)
Similar to *Aug SFI*, this structure addresses the query performance degradation of *VFI*, by augmenting the leaf nodes of the secondary R*-trees with pointers to *I.v*. The space cost with *Aug VFI* is shown in Table 3. Compared with *VFI*, the primary LSH in *Aug VFI* does not need to store pointers to visual data because the hash value of a visual vector *I.v* does not change so it is enough to store both spatial and visual data in the secondary R*-trees. In the case of *Aug SFI*, leaf nodes should maintain pointers to data because of splitting operations which may change the contents of leaf nodes.

When executing a spatial-visual query, the primary LSH uses *Q.v* to identify the buckets that potentially contain similar images to $I_Q$. The secondary R*-trees associated with these candidate buckets are queried using *Q.s*. Augmenting R*-tree leaf nodes with *I.v* enables the validation of the visual similarity of the results after being retrieved from the secondary structures. There is another variation of the query execution in which the secondary R*-trees are explored to retrieve more results. When executing *Q.s.* on each secondary R*-tree, we can enlarge the spatial query by a ratio (referred as spatial exploration ratio $\mathcal{E}.s$). This spatial-explorative query (referred as *Aug VFI-E*), minimizes the effect of image geo-location inaccuracy.

With our running example, when executing *Q.v*, the primary LSH loads two buckets <$B_3$, $B_6$>. Then, *Q.s* is executed on $R^*$-$tree_3$ and $R^*$-$tree_6$. Initially, the list <$I_3$, $I_4$, $I_7$> is reported as the intermediate result but $I_7$ will be discarded because its visual similarity distance from $I_Q$ is greater than $\sigma$ (i.e., $\Phi(I_7.v, I_Q.v) >$ 0.5). Hence, <$I_3$, $I_4$> is reported as the final result.

*Result Accuracy:* The result of *Aug VFI* is identical to *VFI* and *DI*. *Aug VFI–E* can retrieve some of *S-UNMatch-Rel* images based on the exploration ratio.

Accuracy of the results of all discussed index structures is summarized in Table 5.

*Index Performance:* Similar to *VFI*, this structure performs well when the visual selectivity of the query is high. However, the query performance is improved compared to *VFI* because *Aug VFI* reduces the number of I/O operations. As shown in Table 4, the query cost with *Aug VFI* is similar to that with *VFI* but avoids the cost of $T_{LSH}$ part. We further observe the following lemmas:

**Lemma 7:** *QueryIOCost(Aug VFI) ≤ QueryIOCost(VFI)*

*Proof:* Querying *Aug VFI* does not require loading the buckets in the primary LSH because the hashing operation identifies the target buckets and their corresponding secondary R*-trees. Hence, $T_{LSH}^{Aug\ VFI} = O(1)$. In addition, given that for each bucket $b \in [1, B]$, a subset of data objects referenced by $b$ is also referenced by $m_b \in [1, \bar{M}]$. Hence, we can conclude that $m_b \leq C(b)/P_{disk}$ and $\sum_{m=1}^{\bar{M}} P_V(m_b) \leq P_v(b)$. Hence, $T_{Data}^{Aug\ VFI} \leq T_{Data}^{VFI}$.

**Lemma 8:** *QueryIOCost(Aug VFI-E) ≤ QueryIOCost(Aug LSH)*

*Proof:* Derived from Lemma 3 (details omitted due to space limitation).

**Table 5: Result Accuracy of Various Index Structures**

| Index Structure | SV-Match-Rel | S-UNMatch-Rel | V-UNMatch-Rel |
|---|---|---|---|
| *Aug R*-tree* | ✓ | ✗ | ✓ |
| *Aug SFI–E* | ✓ | ✗ | ✓ |
| *Aug VFI–E* | ✓ | ✓ | ✗ |
| Others | ✓ | ✗ | ✗ |

## 4. Experiments
### 4.1 Experimental Methodology
*Dataset:* We used three real world datasets: Flickr (referred as *Flickr*), Google Street View (referred as *GSV*), and *GeoUGV*. *Flickr* contains 185k geo-tagged images in partial areas of Los Angeles, CA; Pittsburgh, PA; and Orlando, FL. *GSV* contains 52k high quality Google Street View images covering the downtown and neighboring areas of Pittsburgh, PA; Orlando, FL and partially Manhattan, NY. Each image in *GSV* is tagged with a GPS location and direction but we used only the GPS location to represent the spatial property of the image. *GeoUGV* is a public geo-tagged user-generated video dataset consisted of 1.6k videos, which were collected by two research prototype systems; MediaQ (http://mediaq.usc.edu/) and GeoVid (http://geovid.org/), recorded at different cities mainly Los Angeles, Singapore and Munich. We processed the video set and extracted a representative frame per second which is tagged with spatial metadata (i.e., Field-of-View, FOV) but we only use a single geo-coordinate (i.e., GPS location) to represent each image. In total, *GeoUGV* contains 124k geo-tagged images. Some statistics of our experimental datasets are shown in Table 6. It is clear that *GSV* is the densest dataset spatially and visually while *Flickr* is the least dense dataset. Note that the size of *GeoUGV*' spatial data file is larger than that of *Flickr* –despite the fact that number of pictures in *GeoUGV* is less than *Flickr's*– because the image ID in GeoUGV is represented by 34 characters while Flickr's image ID has 6 characters.

Each image in these datasets is represented by two vectors: *I.s* and *I.v*. *I.s* is a 2-d vector (latitude and longitude) while *I.v* is a 150-d vector consisted of PCA-CNN image descriptor. Each image was processed using Caffe [35] framework (with default model) to extract 4096-d CNN feature descriptor. Then we applied PCA to reduce the dimensions of the visual vectors (150-d vector).

**Table 6: Dataset Statistics**

| Property | GSV | GeoUGV | Flickr |
|---|---|---|---|
| # of images | 52k | 124k | 185k |
| Size of spatial data | 2mb | 10mb | 4mb |
| Size of visual data | 143mb | 347mb | 509mb |
| # of images in 1*1 km$^2$ | avg.: 1.1k, max: 8k | avg.: 263, max: 30k | avg.: 159, max: 16k |
| # of similar images for ∀ $I_Q$ when $\sigma$=35 | avg.: 63, max: 1.6k | avg.: 55, max: 1.5k | avg.: 37, max: 1.9k |

*Index settings:* We implemented all of our index structures in Java 1.7. All of them are disk resident. We used the page size of 4KB. Our R*-tree had fan-out of 85. Meanwhile, our LSH used 3 hash tables (i.e., *T*=3) with the Euclidean hash family (composed of 7 hash functions (i.e., *F*=7)). The value of *W* in LSH is fixed depending on the dataset (i.e., *W*=100 for *GSV*, *W*=95 for *GeoUGV*, and *W*=90 for *Flickr*). The spatial and visual vectors are stored in a plain text file, and R*-tree leaf nodes and LSH buckets store pointers to the vectors' locations in the file. All experiments were performed on a 3.6 GHz Intel Core i7 machine with 12 GB memory and 1TB 7200RPM disk drive running on 64 bit Windows 7.

*Queries and Metrics:* Table 7 lists all query settings with the default values underlined. In our experiments, we need to construct queries with different spatial and visual selectivity factors. Hence, we need to select a query image that generates the $I_Q.v$ and $I_Q.s$ vectors with desired selectivity. Towards this end, we first partitioned each dataset spatially and visually, separately, based on their density (i.e., dense, uniform, sparse). Next, we merged them to generate five groups: spatial dense visual dense (*SD-VD*), spatial dense visual sparse (*SD-VS*), spatial sparse visual dense (*SS-VD*), spatial sparse visual sparse (*SS-VS*), and spatial uniform visual uniform (*SU-VU*). We refer to these groups as *Query Selectivity*. For each query selectivity group, we randomly selected 250 query images $I_Q$ (e.g., a query image selected from *SD-VS* would generate a query with low spatial selectivity and high visual selectivity. Hence, we used *SU-VU* as the default selectivity to avoid selectivity impact). In addition, the selected spatial ranges (ascendingly ordered in Table 7) represent the campus areas of four universities: University of Southern California, University of Wisconsin – Madison, University of California – Irvine, and University of Florida, respectively. To evaluate the index structures, we report two metrics: a) result accuracy in terms of recall[4] when executing $Q_{range}$, and b) index performance in terms of the number of accessed pages and query time. To estimate ground truth, in order to calculate recall, we merged all relevant images retrieved from *Aug R*-tree* and *Aug VFI-E* for each $Q_{range}$. The query time is calculated to measure two primary parts of the query execution: I/O cost when loading pages from disk and index overhead. The index overhead is mainly represented by three parts: index traversal (e.g., traversing R*-tree to locate target leaf nodes or hashing operations in LSH to locate target buckets), data filtering (e.g., evaluate intermediate results by Euclidean distance calculations), and merging data (e.g., combining intermediate results of LSH and R*-tree in *DI*). To avoid caching effect on timing, each query was executed 5 times and the longest and shortest times were ignored and the query time is the average of the remaining.

**Table 7: Query Settings**

| Query Parameter | Values |
|---|---|
| Query Selectivity | SD-VD, SD-VS, SS-VD, SS-VS, **SU-VU** |
| Spatial Range | 1.25*1.25 km$^2$, 3.7*3.7 km$^2$, **6.18*6.18 km$^2$**, 8.1*8.1 km$^2$ |
| Visual Range | 25, 30, **35**, 40 |
| Spatial Exploration Ratio | 0.1, 0.3, **0.5**, 0.7 |
| Visual Exploration Ratio | 9, **15**, 21, 27 |

### 4.2 Index Construction:

*Space Cost:* Based on our space cost model, the space is represented by two factors: size of index entities (e.g., node or bucket files) and size of data. Because the cost of data files are the same across all index structures for each dataset, we report only the space cost based on index size (see Fig. 6). Among the baseline structures, *DI* is the largest because it combines two separate index structures. Note that the size of R*-tree index is smaller than that of LSH due to two reasons: 1) the R*-tree structure minimizes the depth of the tree and the number of nodes, under the constraint of tree fan-out while the LSH structure does not have any constraint on the size and number of generated buckets, and 2) R*-tree inserts the spatial vector of an image once while LSH insert the visual vector multiple times based on the number of Hash Tables. Consequently, the size of *Aug R*-tree* is smaller than that of *Aug LSH*. Furthermore, *VFI* consists of multiple R*-trees while *SFI* consists of multiple LSHs; thus the size of *VFI* is smaller than that of *SFI*. The same observation holds for the augmented[5] hybrid index structures with a little more space requirement with *Aug SFI* since it stores additional pointers in the secondary LSHs but less space with *Aug VFI* since it does not store pointers in the primary structure.

*Insertion Time Cost:* For each index structure, we report the average time of inserting a geo-tagged image as shown in Fig. 7. Among the baselines, *DI* takes the longest time because it maintains two individual index structures. R*-tree takes longer time than LSH because insertion of an object in R*-tree might require re-organizing nodes by splitting operations which costs additional time. Meanwhile, inserting an object in LSH requires

---

[4] The precision of all index structures is 100% which means that these structures do not retrieve irrelevant images to $I_Q$.

[5] Space cost with *Aug SFI* and *Aug VFI* are omitted from Fig. 6 to avoid crowding the graph

only hashing operations and appending the objects into the target buckets. Consequently, the average insertion time with *Aug R\*-tree* is greater than that of *Aug LSH*. Moreover, among the hybrid index structures *SFI* has the highest insertion time due to maintain one large R\*-tree. Insertion time with *SFI* is almost similar to the one with *DI* because the insertion cost in smaller LSH (in case of *SFI*) or larger LSH (in case of *DI*) are almost the same. Meanwhile, insertion time with *VFI* is smaller than the one with *DI* (with a speedup factor 3.6 in *GSV*, 2.5 in *GeoUGV* and 2.3 in *Flickr*) because maintaining small R\*-trees requires less time.

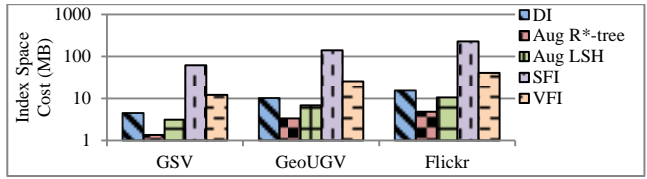

**Figure 6: Size of Various Indexes across Different Datasets**

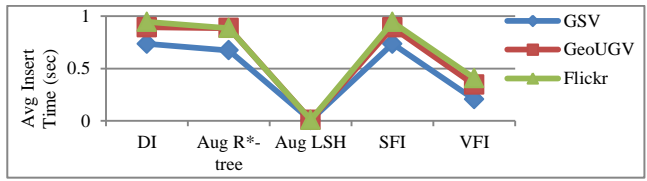

**Figure 7: Avg. Insertion Time for Indexes w/ All Datasets**

## 4.3 Result Accuracy

***Baseline vs. Hybrid:*** In this set of experiments, we used the default query settings to evaluate the recall of the baseline compared to the hybrid structures across different datasets. Even though our index structures achieve 100% precision, they might observe lower recall due to the failure of retrieving all relevant images. As shown in Fig. 8, among the baseline structures, *Aug R\*-tree* achieved the highest recall since it avoids the inaccuracy caused by LSH. The recall of the other baselines and the hybrid structures[6] is identical because they retrieve the same class of images (i.e., *SV-Match-Rel*).

***Impact of Query Selectivity on Hybrid Structures:*** In this set of experiments, we varied the query selectivity parameter to evaluate its effect on the recall of different hybrid structures using *Flickr*[7]. In Fig. 9, all hybrid index structures had identical recall for a given query selectivity; however, the recall value varied across various query selectivity factors. In particular, all structures observed the worst recall value (51%) with *SD-VD* while they achieved the best recall (81%) with *SS-VD*. With *SD-VD,* there are many images that are spatially crowded and visually similar; thus, the probability of having partially similar images is high. This increases the LSH inaccuracy and hence incurs low recall. Meanwhile, with *SS-VD* there are a few images but they are very similar to each other which decreases the LSH inaccuracy; hence achieving high recall. In Fig. 10, we show the effect of explorative technique on the result accuracy when varying the query selectivity. In general we gained the higher recall improvement when exploring visually (i.e., *Aug SFI-E*) compared with spatial exploration (i.e., *Aug VFI-E*). The impact of visual exploration with *Aug SFI-E* was the highest with *SD-VD* increasing the recall of *Aug SFI* by roughly 50%. As mentioned earlier, with *SD-VD* the LSH inaccuracy increases and this deficiency can be minimized by *Aug SFI-E* which retrieves some of *V-UNMatch-Rel* images. Meanwhile, the impact of spatial exploration with *Aug VFI-E* was the highest with *SS-VD* with only 5% improvement. With *SS-VD*, there are a few images nearby $I_Q$ but the number of similar images is potentially high. Hence, exploring spatially with *SS-VD* might result in more images (i.e., *S-UNMatch-Rel*) that are relevant to $I_Q$.

---

[6] The recall of hybrid structures without the effect of exploration is the same. Later, we will show the effect of exploration.

[7] Hereafter, we report the results only for Flickr as the same trends hold for the other datasets.

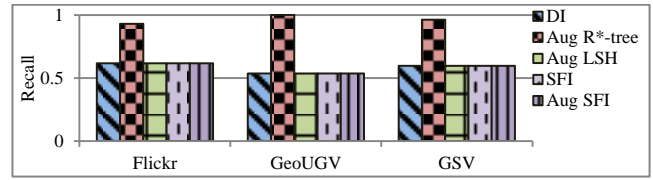

**Figure 8: Recall of Baseline vs. Hybrid**

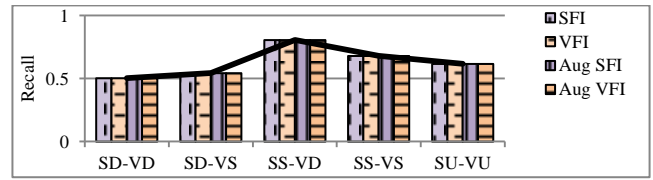

**Figure 9: Impact of Query Selectivity on Hybrid w/ *Flickr***

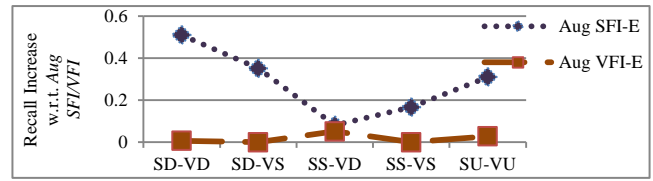

**Figure 10: Impact of Exploration on Recall w/ *Flickr***

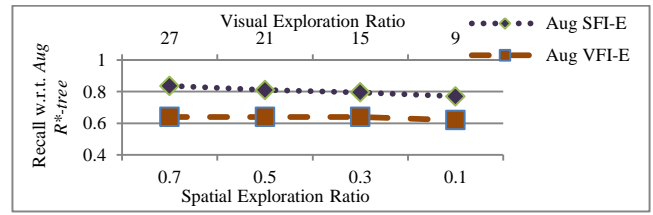

**Figure 11: Impact of Exploring Spatially (Bottom X-Axis) and Visually (Top X-axis) on Recall w/ *Flickr***

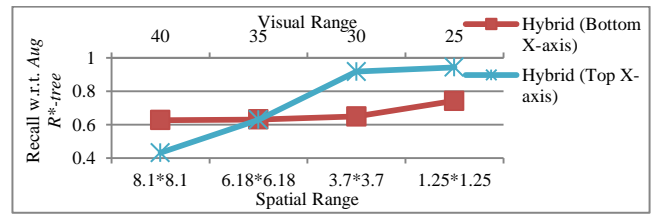

**Figure 12: Impact of Spatial Range (Bottom X-Axis) and Visual Range (Top X-axis) on Recall w/ *Flickr***

***The Impact of Exploration Ratio:*** In this set of experiments (see Fig. 11), we varied the visual exploration ratio $\mathcal{E}_v$ (i.e., the top X-axis) or spatial exploration ratio $\mathcal{E}_s$ (i.e., the bottom X-axis) to evaluate their effect on the recall of hybrid structures with respect to *Aug R\*-tree* using *Flickr*. We chose *Aug R\*-tree* as the benchmark because it showed the best recall. As shown in Fig. 11, increasing the visual exploration ratio improved the recall considerably while increasing the spatial exploration ratio did not show continuous improvement because the missed *S-UNMatch-*

*Rel* images might be located nearby the original spatial range, hence exploring spatially within the nearby vicinity might be sufficient to retrieve them. In particular, the recall of *Aug SFI-E* (i.e., visual exploration) reached 84% and 77% of that with *Aug R\*-tree* when $\mathcal{E}.v = 27$ and $\mathcal{E}.v = 9$, respectively. Meanwhile, the recall of *Aug VFI-E* (i.e., spatial exploration) reached 64% and 62% of that with *Aug R\*-tree* when $\mathcal{E}.s = 0.9$ and $\mathcal{E}.s = 0.1$, respectively.

*The Impact of Spatial Range:* In this set of experiments (see Fig. 12), we varied the spatial range (i.e., the bottom X-axis) to evaluate its effect on the recall of hybrid structures with respect to *Aug R\*-tree* using *Flickr*. As shown in Fig. 12 when shrinking the spatial range, the recall of hybrid structures (i.e., *SFI, VFI, Aug SFI,* or *Aug VFI*) slightly improved approaching the recall of *Aug R\*-tree*. Decreasing the spatial range restricts the search space to the images which are spatially very close and potentially very similar to $I_Q$; hence decreasing LSH inaccuracy. This leads to less retrieval of *V-UNMatch-Rel* images by *Aug R\*-tree* in favor of retrieving more *SV-Match-Rel* images.

*The Impact of Visual Range:* In this set of experiments (see Fig. 12), we varied the visual range (i.e., the top X-axis) to evaluate its effect on the recall of hybrid structures with respect to *Aug R\*-tree* using *Flickr*. As shown in Fig. 12, when shrinking the visual range, the recall of hybrid structures (i.e., *SFI, VFI, Aug SFI,* and *Aug VFI*) considerably improved approaching the recall of *Aug R\*-tree*. Decreasing the visual range restricts the search space only to very similar images that are potentially not affected with the inaccuracy of LSH leading to less retrieval of *V-UNMatch-Rel* images by *Aug R\*-tree* in favor of retrieving more *SV-Match-Rel* images.

## 4.4 Index Performance

*Baseline vs. Hybrid:* In this set of experiments, we used the default query settings to evaluate the performance of the baseline compared to the hybrid structures[8] across different datasets. Fig. 13 depicts the results, where the Y-axis is the number of pages accessed in logarithmic scale. As shown in Fig. 13, the baseline structures (*DI, Aug R\*-tree,* and *Aug LSH*) performed worse than the hybrid index structures for all datasets. In particular, *Aug R\*-tree* across all datasets incurred the worst performance. Compared to *DI* and *Aug LSH*, *Aug R\*-tree* suffers from lower performance due to two reasons: 1) the large size of the augmented visual vectors, and 2) potential retrieval of additional images (i.e., *V-UNMatch-Rel*). Meanwhile, the hybrid structures were superior because they organize data in a two-level structure which enables pruning the search space for the secondary structures. Furthermore, the augmented hybrid structures provided additional speedup because the primary structure does not retrieve intermediate results. Furthermore, we considered index overhead time added to I/O cost and the same observations still hold as shown in Fig. 14. Note that the visual-first hybrid structures follow the same trend as compared to the baselines.

*Impact of Query Selectivity on Hybrid Structures:* In this set of experiments, we varied the query selectivity parameter to evaluate its effect on the performance of different hybrid structures with *Flickr*[9]. As shown in Fig. 15 (the Y-axis is the number of pages accessed in logarithmic scale), the performance of *SFI* and *VFI* varied inversely. In particular, *VFI* outperformed *SFI* with speedup factor of 2.14 with the query selectivity *SD-VS*, because the spatial selectivity factor is lower than the visual one. With the other types of query selectivity, *SFI* was superior achieving the best speedup factor of 3.13 with *SS-VD* because the spatial selectivity is the highest. Furthermore, the relation between *Aug SFI* and *Aug VFI* is analogous to that of *SFI* with *VFI* but with better performance. *Aug SFI* achieved the best speedup with respect to *SFI* by a factor of 5.5 with *SS-VS* while the best speedup factor of *Aug VFI* with respect to *VFI* was 7.4 with *SS-VS*. The query time comparison (in seconds) of the hybrid structures shown in Fig. 16 conveys the same observations obtained in Fig. 15. In Fig. 17, we show the effect of explorative technique on the performance when varying the query selectivity. In general exploring visually (i.e., *Aug SFI-E*) causes higher slow-down than exploring spatially (i.e., *Aug VFI-E*) because *Aug SFI-E* explores in the secondary LSHs whose buckets don't have a limit on their size while *Aug VFI-E* explores the secondary R\*-tree whose leaf nodes have limited size. *Aug SFI-E* showed the least slow-down factor with *SD-VD*, which showed the highest recall improvement, because there are many relative images with high probability being scattered in few disjointed buckets. Moreover, with *SD-VD Aug VFI-E* showed the least slow-down factor because the potential similar images within the explored spatial ration have a higher probability to be stored to least number of leaf nodes.

*The Impact of Exploration Ratio:* In this set of experiments (see Fig. 18), we varied the visual exploration ratio $\mathcal{E}.v$ (i.e., the top X-axis) or spatial exploration ratio $\mathcal{E}.s$ (i.e., the bottom X-axis) to evaluate their effect on the performance[10] of hybrid structures with respect to *Aug R\*-tree* using *Flickr*. We chose *Aug R\*-tree* as the benchmark because it showed the worst performance along with the best recall. As shown in Fig. 18, the performance degradation incurred by the spatial exploration (i.e., *Aug VFI-E*) is less than the performance degradation incurred by the visual exploration. In particular, the performance speedup of *Aug SFI-E* (i.e., visual exploration) reached 85.6% and 87.9% when $\mathcal{E}.v = 27$ and $\mathcal{E}.v = 9$, respectively. Meanwhile, the performance speedup of *Aug VFI-E* (i.e., spatial exploration) reached 96.3% and 96.6% when $\mathcal{E}.s = 0.9$ and $\mathcal{E}.s = 0.1$, respectively.

*The Impact of Spatial Range:* In this set of experiments, we varied the spatial range to evaluate its effect on the performance of hybrid structures with respect to *Aug R\*-tree* using *Flickr*. As shown in Fig. 19, the speedups of both *VFI* and *Aug VFI* were directly affected when shrinking the spatial range while the speedups of both *SFI* and *Aug SFI* were almost steady (with minor improvement). When decreasing the spatial range, *SFI* saves many disk accesses in the primary structure (i.e., R\*-tree) similarly to *Aug R\*-tree*; hence the speedup of *SFI* was marginal. Meanwhile, *VFI* does not save many disk accesses since its primary index (i.e., LSH) is not affected when varying the spatial range. In particular, *VFI* only saves a few disk accesses with its spatial secondary indexes.

---

[8] We only used our spatial-first variations as representatives for hybrid to avoid crowding the graph. Later, we include more in-depth comparisons between spatial-first and visual-first by varying the query selectivity.

[9] Hereafter, we report the results only for *Flickr* as the same trends hold for the other datasets.

[10] Hereafter, we report the performance in terms of I/O cost due to space limitation.

Varying the visual range does not affect the performance of our index structures in terms of disk accesses because the similarity filter is applied after retrieving all visual vectors from the disk.

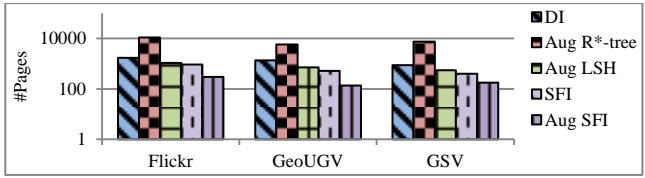

Figure 13: Performance (I/O Cost) of Baseline vs. Hybrid

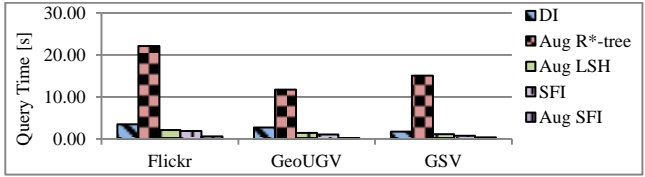

Figure 14: Performance (Query Time) of Baseline vs. Hybrid

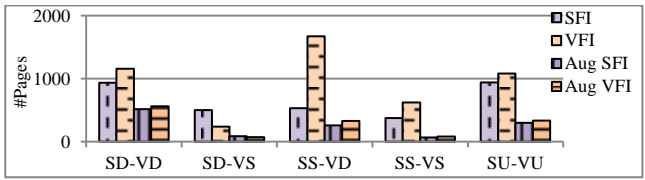

Figure 15: Impact of Query Selectivity on Hybrid w/ *Flickr*

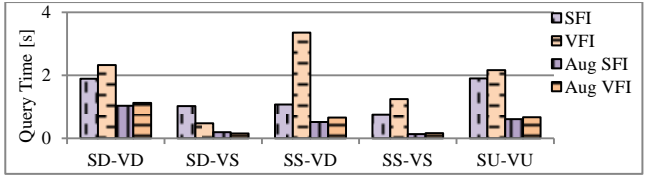

Figure 16: Impact of Query Selectivity on Hybrid w/ Flickr

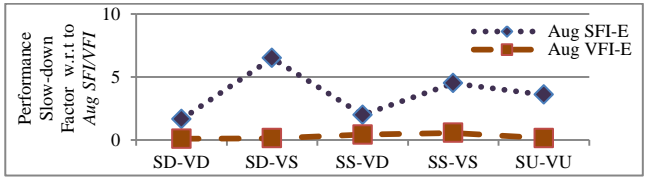

Figure 17: Impact of Exploration on Performance w/ *Flickr*

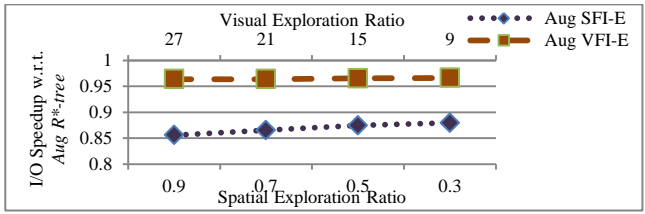

Figure 18: Impact of Exploring Spatially/Visually on Performance w/ Flickr

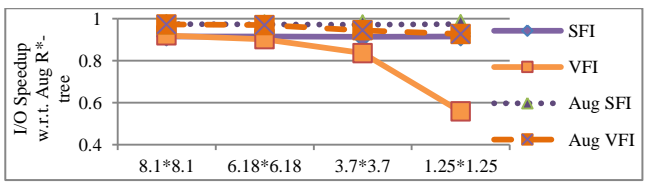

Figure 19: Impact of Spatial Range on Hybrid w/ Flickr

In summary, the hybrid structures outperformed the baselines but choosing the best hybrid structure mainly depends on the query selectivity. Among the variations of *SFI*, *Aug SFI-E* was superior in terms of recall while *Aug SFI* achieved the best performance. Meanwhile, among *VFI*'s variations, *Aug VFI-E* carried out a marginal recall increase but *Aug VFI* scored the best performance.

## 5. RELATED WORK

*Spatial-Textual Indexing:* Initially the focus of the database research community was on designing index structures for a single data type. Due to the evolution of sensor-rich devices and location-based applications, the focus was switched to multi-type index structures. One active research area in multi-type indexing is spatial-textual search. The keyword inverted index file is the main structure that has been utilized in spatial-textual indexing. One group of research studies which focused on utilizing R*-tree for spatial-textual indexing includes: I) Two structures introduced in [11] are First Inverted File Then R*-tree and First R*-tree Then Inverted File, II) KR*-tree proposed in [20], which extends First R*-tree Then Inverted File structure by augmenting each node of R*-tree with a list of all keywords that appear in its subtree, and III) $IR^2$-tree presented in [10], which extends R-tree by augmenting each node in the tree with a signature representing the union of the keywords of its subtree. Another group of research studies which utilized Grid for spatial-textual indexing includes: I) Two designs presented in [21] which are similar to the structures in [11] but R*-tree is replaced with grid, and II) Spatial-Keyword Inverted File structure proposed in [13] which combines keywords and grid cells in an inverted index file.

*Spatial-Visual Search:* To the best of our knowledge there are only two studies [8,9] that tackled the spatial-visual search. In [8], the authors proposed only one index structure similar to our *Double Index*. However, their study mainly focused on the spatial-visual ranking algorithm to evaluate kNN query. In [9], a location-sensitive image advertisement platform was envisioned for a real-world advertising system in Beijing. The authors proposed an index structure that is analogous to our *SFI*. Therefore, the focus of both studies was on neither a thorough exploration of the indexing challenges in spatial-visual search nor the comparative evaluation of the result accuracy and performances of various indexes. Finally, both studies utilize the vocabulary tree as visual index structure but alternatively we chose to use LSH as visual index due to the following two reasons. First, LSH is more suitable for indexing global image descriptors (e.g., CNN) while vocabulary tree is built to index local image descriptors (e.g., SIFT). Second, LSH shows better search performance because of using the hashing technique while the vocabulary tree uses the recursive clustering to partition the space resulting in worse performance and higher inaccuracy especially when the tree becomes deeper [22].

## 6. Conclusion

In this paper, we studied the indexing challenges of images with geo-tagged data to expedite spatial-visual search. We proposed a set of hybrid index structures and evaluated them in comparison with a set of baselines in terms of performance and result accuracy of spatial-visual range query. We showed experimentally that all hybrid structures outperformed the baselines with a maximum speed-up factor of 42.7. When shrinking the visual range, our hybrid structures achieved 100% recall; however, the recall was low (53%) for large visual ranges. The visual exploration with the hybrid structures provided 33% recall improvement while still outperforming the highest-recall baseline with a speedup factor of 8.3. In addition, when shrinking the spatial range, the recall of both baseline and hybrid structures decreases. The hybrid structure with spatial exploration was able to improve the recall marginally (by 5%) while still outperforming the highest-recall baseline with a speedup factor of 29.7. For future work, we plan to extend our hybrid index structures to include the direction of the viewable scene in the spatial metadata of the image.